\documentclass[twocolumn,aps,prl,showpacs,amsmath,superscriptaddress,longbibliography,notitlepage]{revtex4-1}
\usepackage{amssymb}
\usepackage{mathrsfs}
\usepackage{graphicx}
\usepackage{float}
\usepackage[caption=false]{subfig}
\usepackage{epstopdf}

\usepackage[normalem]{ulem}
\usepackage{verbatim}
\usepackage{xcolor}
\usepackage{bm}

\def\blue{\textcolor{blue}}
\def\red{\textcolor{red}}

\begin{document}

\def\qv{\vec{q}}
\def\red{\textcolor{red}}
\def\blue{\textcolor{blue}}
\def\magenta{\textcolor{magenta}}
\def\apricot{\textcolor{Apricot}}

\def\GJ{\textcolor{blue}}
\def\LH{\textcolor{Orange}}
\def\TT{\textcolor{ForestGreen}}

\newcommand{\norm}[1]{\left\lVert#1\right\rVert}
\newcommand{\ad}[1]{\text{ad}_{S_{#1}(t)}}

\title{Impurity induced scale-free localization}

\author{Linhu Li}
\affiliation{Department of Physics, National University of Singapore, Singapore, 117542.}
\author{Ching Hua Lee}
\email{phylch@nus.edu.sg}
\affiliation{Department of Physics, National University of Singapore, Singapore, 117542.}
\author{Jiangbin Gong}
\email{phygj@nus.edu.sg}
\affiliation{Department of Physics, National University of Singapore, Singapore, 117542.}
\begin{abstract}
This work develops a full framework for non-Hermitian impurity physics in a non-reciprocal lattice, with PBCs, OBCs and even their interpolations being special cases across a whole range of boundary impurity strengths.  As the impurity strength is tuned,   the localization of steady states can assume very rich behavior, including the expected non-Hermitian skin effect, 
Bloch-like states albeit broken translational invariance, and surprisingly, scale-free accumulation along or even against the direction of non-reciprocity.   We further uncover the possibility of the co-existence of non-Hermitian skin effect and scale-free localization, where qualitative aspects of the system's spectrum can be extremely sensitive to impurity strength. We have also proposed specific circuit setups for experimental detection of the scale-free accumulation, with simulation results confirming our main findings.
\end{abstract}



\date{\today}
%


\maketitle
Spatial inhomogeneity in physical systems is the norm rather than the exception.  It can trigger a wide variety of physical phenomena, such as the Anderson localization, topological edge states and  topological defect states.
In non-Hermitian systems, intriguing physics from spatial inhomogeneity encompasses not just the non-Hermitian skin effect (NHSE)\cite{yao2018edge,yokomizo2019non,Lee2019anatomy,lee2018tidal,kunst2019non,edvardsson2019non,yang2019auxiliary,zhang2019correspondence,brandenbourger2019non,Lee2019hybrid,mu2019emergent,li2019geometric,lee2020ultrafast,longhi2020non,lee2020many,cao2020non,xue2020non,liu2020helical,rosa2020dynamics,yoshida2020mirror,yi2020non,xiao2020non,li2020topological,schomerus2020nonreciprocal,PhysRevLett.124.086801,koch2020bulk,teo2020topological,li2020critical}, but also 
impurity- or defect-induced topological bound states~\cite{bosch2019non,liu2019topological,zhao2020topological,liu2020diagnosis}, 
disorder-driven non-Hermitian topological phase transitions~\cite{luo2019non},
as well as non-Hermitian quasi-crystals and mobility edges with an incommensurate modulation~\cite{longhi2019topological,jiang2019interplay,zeng2020topological,claes2020skin}.  

Due to their emergent non-locality, non-reciprocal impurities in non-Hermitian systems generate dramatic spectral flows as their strengths are varied~\cite{xiong2018does,Lee2019anatomy}. This has even been proposed for exponentially enhanced quantum sensing in an experimentally realistic setting~\cite{budich2020sensor,mcdonald2020exponentially}. Yet, there does not exist a full framework for non-Hermitian impurity physics, with periodic and open boundary conditions (PBCs and OBCs) being special cases across a whole range of boundary impurity strengths. This work aims to fill in this important gap and reports unexpected findings of general theoretical and experimental interest. 

Specifically, we discover that boundary impurities in non-reciprocal lattices can generate new types of steady-state localization behavior characterized by scale-free accumulation (SFA) of eigenstates, despite having non-power-law profile. In sharp contrast to the NHSE, the SFA direction can be counter-intuitive, opposite of the non-reciprocal directionality. With varying impurity strengths, the steady state makes transitions between 
the NHSE behavior, Bloch-like eigenstates with broken translational invariance, ordinary SFA,  and reversed SFA. 
A careful inspection of these qualitatively rich transitions reveals fascinating duality relations between weak and strong inhomogeneity, yielding a big picture of non-Hermitian impurity physics. Known NHSE properties are thus revealed as only one of the many impurity-induced consequences in non-reciprocal non-Hermitian systems. Drastically different steady-state behaviors can even co-exist when next nearest hoppings are present, a useful phenomenon that can benchmark the hyper-sensitivity of non-Hermitian systems to boundary/impurity effects.  
   
\begin{figure}
\includegraphics[width=1\linewidth]{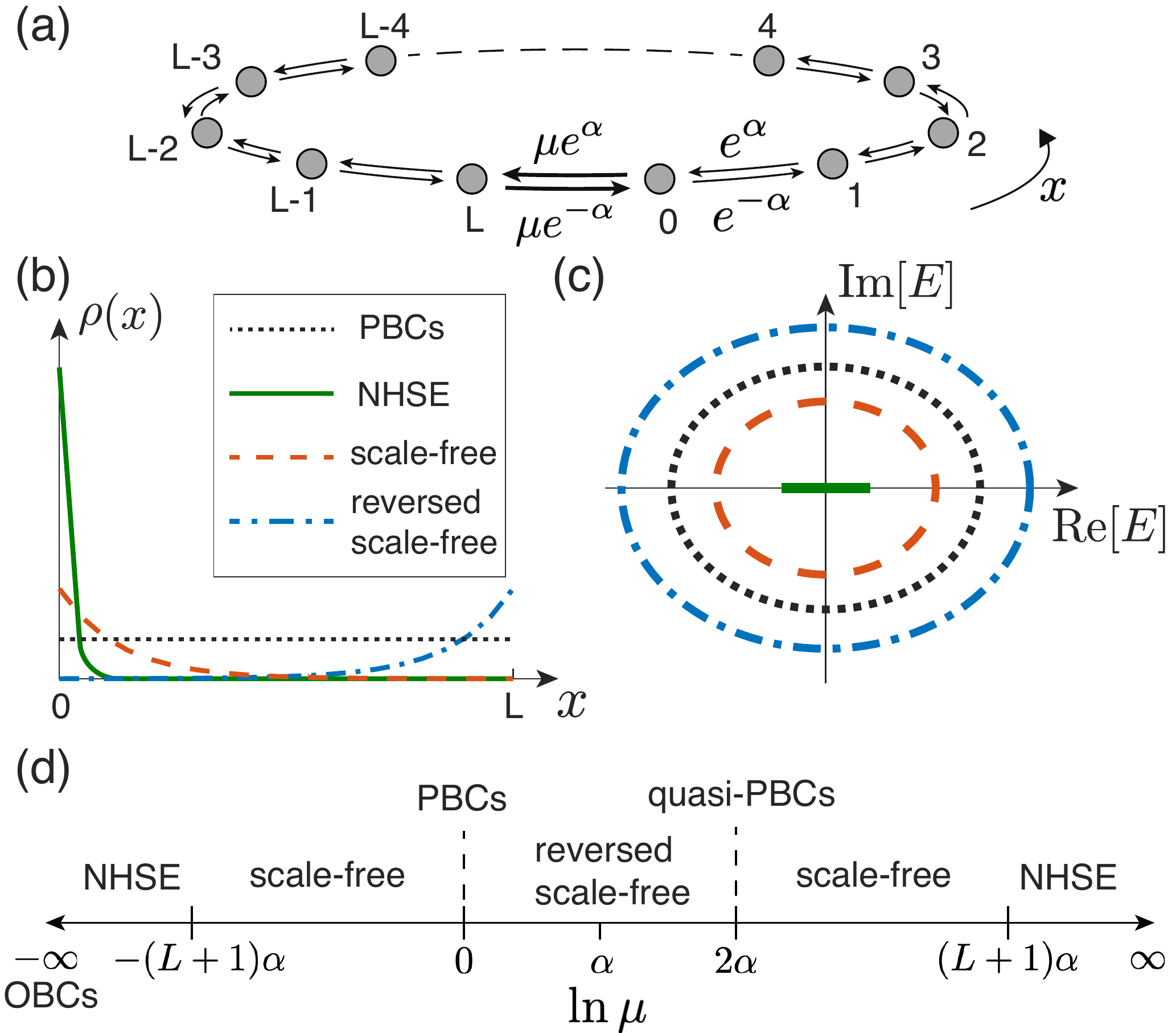}
\caption{(a) The Hatano-Nelson model with impurity couplings $\mu e^{\pm\alpha}$. PBCs and OBCs correspond to $\mu=1$ and $0$ respectively, but other $\mu$ support qualitatively different phenomena. 
(b) Spatial eigenstate distributions $\rho(x)=|\psi_x|^2$ with different accumulating behaviors, with $\psi_x$ the wave-function value at $x$.
(c) Complex spectra distinguishing the four types of eigenstates in (b).
(d) Different regimes across the whole range $\mu$ are marked by different accumulation phenomena, with dualities relating strong and weak $\mu$.
}
\label{fig:sketch}
\end{figure}

\noindent{\it Impurity-induced SFA. -- }
We consider impurities in the simplest 1D Hatano-Nelson chain~\cite{HN1996prl}, which already exhibits nearly the full scope of impurity-induced phenomena in more generic lattices. An impurity is represented as a modified coupling between the first and last sites:
\begin{eqnarray}
H=\sum_{x=0}^{L-1} \left[e^{\alpha}\hat{c}^\dagger_x\hat{c}_{x+1}+e^{-\alpha}\hat{c}^\dagger_x\hat{c}_{x-1}\right]+\mu_+\hat{c}^\dagger_L\hat{c}_{0}+\mu_-\hat{c}^\dagger_0\hat{c}_{L}\nonumber
\end{eqnarray}
with $\mu_{\pm}=\mu e^{\pm\alpha}$, 
$\mu$ controlling the local impurity,
$\alpha>0$ and $x=0,1,...L$ labeling the lattice sites [Fig.~\ref{fig:sketch}(a)]. 
PBCs are recovered at $\mu=1$, where translational symmetry is restored and the system can be described by a Bloch Hamiltonian $H(z)=e^{\alpha}z+e^{-\alpha_/z}$ with $z=e^{ik}$, $k$ the quasi-momentum.
Perfect OBCs yielding the NHSE are recovered at $\mu=0$, although a finite-size system behaves like OBCs when $\mu\lesssim e^{-\alpha(L+1)}$~\cite{koch2020bulk}.  Cases with $0<\mu<1$ may be interpreted as interpolations between PBCs and OBCs, but a full picture with new physics emerges only if the whole range of $0<\mu<\infty$ is investigated.
Beyond $\mu\in [0,1]$, eigenstates can exhibit weaker boundary accumulation toward either direction, even unexpectedly against the direction of non-reciprocity and NHSE [Fig.~\ref{fig:sketch}(b)]. 
Furthermore, this intriguing localization phenomenon is dubbed the SFA because the eigenstates display a scale-free spatial profile, decaying as $e^{-Cx/L}$ with constant $C$, as elaborated later. 
Unlike in the NHSE, the spectrum of these SFA eigenstates forms a loop that can be deformed away from, or even enclosing the PBC spectrum [Fig.~\ref{fig:sketch}(c)].
{Different accumulation regimes exist for $\mu$ ranging from $0$ to $\infty$, and notably similar behaviors are seen in both the small and large $\mu$ limits [Fig.~\ref{fig:sketch}(d)]. As detailed in our concrete examples later, we find two types of dualities between $\mu$ and $\sim 1/\mu$, 
which allow us to probe the large $\mu$ regime from the small $\mu$ regime, and vice versa.}

\noindent{\it {Ordinary} and reversed SFA. -- }
To understand why SFA occurs, we analytically solve for the eigenstates $\Psi_n=\sum_x^L\psi_{x,n}\hat{c}^\dagger_x|0\rangle$, $n=0,...,L$ via $H\Psi_n=E_n\Psi_n$, under reasonable approximations. 
In the large-$\mu$ limit with $\mu \gg e^{\pm\alpha}$,  two isolated eigenstates strongly localize at $x=0,L$, with eigenenergies $E_{\rm iso}\approx\pm\mu$~\cite{SuppMat}.  
The other eigenstates are exponentially decaying:
\begin{eqnarray}
\psi_{x,n}=e^{-\left[\kappa_L-i\frac{(2n+1)\pi}{L-1}\right](x-1)},x\neq0,\kappa_L=\frac{\ln\mu -2\alpha}{L-1}\label{eq:dis_hopping}
\end{eqnarray}
with $\psi_{0,n}\approx0$~\cite{SuppMat}, $n=1,2,...,L-1$ yielding $L-1$ different eigenstates. 
Physically, the vanishing amplitude at $x=0$ can be partially appreciated by the physics underlying electromagnetic field induced transparency~\cite{EIT}.  That is, the much stronger coupling between sites $L$ and $0$
effectively makes the rest of the lattice
more ``transparent'', and hence suppresses the population pumping from the rest of the lattice to site 0~\cite{SuppMat}. 
In a more restricted parameter regime with $e^\alpha\gg e^{-\alpha}$ and $\mu\ll e^{\alpha(L+1)}$, the corresponding eigenenergies can be further approximated by
\begin{eqnarray}
E_n
\approx\epsilon(k_n+i\kappa_L)\label{eq:E_hopping}
\end{eqnarray}
with $k_n:=(2n+1)\pi/(L-1)$ and $\epsilon(k)$ the eigenenergy function at $\mu=1$ (i.e. PBCs)~\cite{SuppMat}.
Remarkably, the spectrum is obtainable via a complex deformation of the PBC quasi-momentum, similar to the GBZ approach for OBC systems~\cite{yao2018edge,yokomizo2019non,Lee2019anatomy}.
Yet, the $1/(L-1)$ coefficient in the decay exponent $\kappa_L$ indicates much weaker accumulation for a large system, and in fact suggests a {\it scale-free} decay profile from $x=1$ to $L$.
The dependence of $\kappa_L$ on $\mu$ (and $L$) differs from that of impurity-induced topological localization~\cite{SuppMat}.

\begin{figure}
\includegraphics[width=1\linewidth]{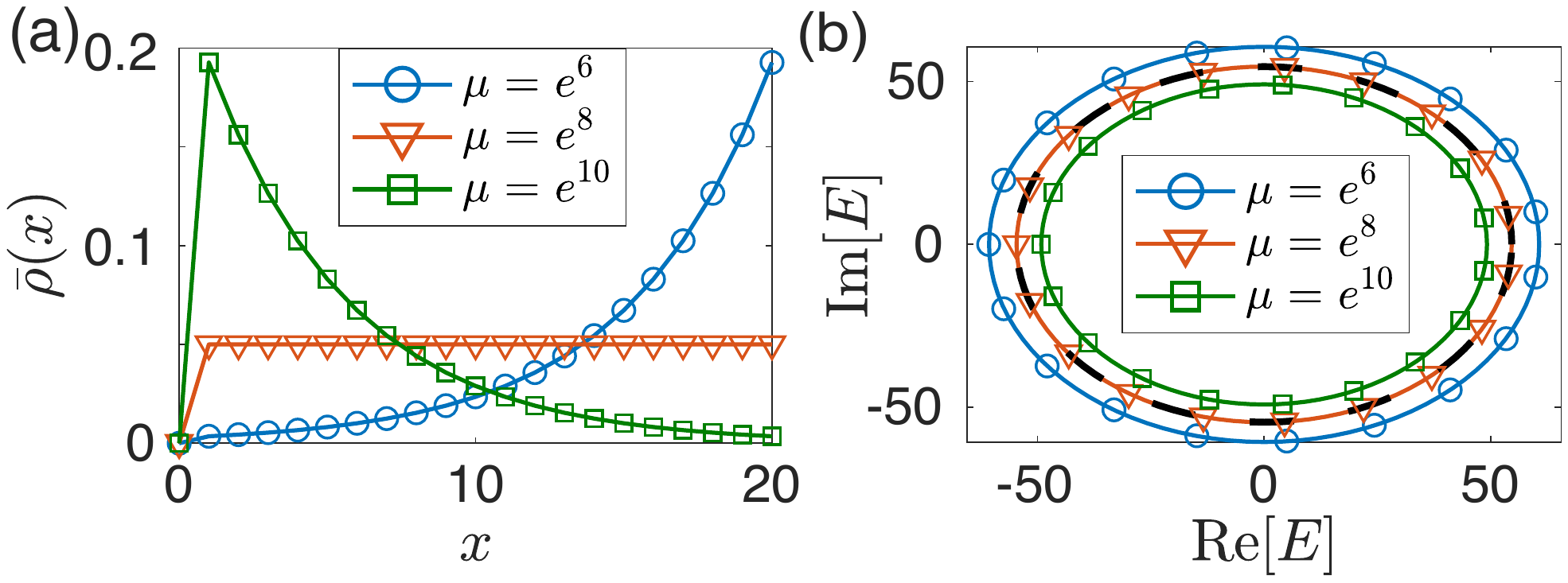}
\caption{(a) Average distribution of all the anomalously accumulating eigenstates $\bar{\rho}(x)=\sum_{n=1}^{L-1} \rho_n(x)/(L-1)$ with $\rho_n(x)=|\psi_{x,n}|^2$ and $\alpha=4$, $L=20$. The blue, red and green cases correspond to reversed SFA, quasi-PBCs and SFA. (b) Spectra for these anomalously accumulating eigenstates, excluding the two isolated eigenstates. The black dashed curve is the true PBC spectrum ($\mu=1$), which overlaps with the quasi-PBC red curve at $\mu=e^{2\alpha}=e^8$. In both panels, the circles, squares, and triangles are numerical data points, and the colored solid curves are approximations from Eqs.~(\ref{eq:dis_hopping}),(\ref{eq:E_hopping}).
}
\label{fig:approx}
\end{figure}

Counter-intuitively, reversed accumulation with negative $\kappa_L$ can occur when $\mu<e^{2\alpha}$, which still falls into a valid sub-regime if $\mu \gg e^{\alpha} \gg e^{-\alpha}$, as confirmed by the agreement between our approximate solutions and numerical results in Fig.~\ref{fig:approx}(a). 
For the peculiar borderline case of  $\mu=e^{2\alpha}$ between ordinary and reversed SFA, $\kappa_L=0$ and the eigenstates are uniformly distributed (except at $x=0$) and hence resemble Bloch states [Eq.~\ref{eq:dis_hopping} and Fig.~\ref{fig:approx}(a)], even though translational invariance is broken. Indeed, the continuous part of the associated spectrum also coincides with the PBC spectrum [Eq.~\ref{eq:E_hopping} and Fig~\ref{fig:approx}(b)]. This curious case of quasi-PBC delocalized states is elaborated in the Supplemental Materials~\cite{SuppMat}.
While we have considered a strong non-reciprocity of $\alpha=4$ in Fig.~\ref{fig:approx} for a better illustration, more examples with weaker $\alpha$ are found in~\cite{SuppMat}.


\noindent{\it Duality between strong and weak impurity couplings.-- }
the discussions above imply a duality between PBCs at $\mu=1$ and quasi-PBCs at $\mu=e^{2\alpha}$. 
This motivates us to seek duality relations for the whole range of $\mu$.  
For $e^{-2\alpha}\ll\mu\ll1$,
another set of exponentially decaying eigenfunctions are found, i.e.,
\begin{eqnarray}
\psi'_{x,n}=e^{-\left[\kappa_L'-i\frac{2n\pi}{L+1}\right]x},\kappa_L'=\frac{-\ln\mu }{L+1},\label{eq:dis_hopping_weak}
\end{eqnarray}
with
\begin{eqnarray}
E'_n\approx e^\alpha e^{\left[\ln \mu+i2n\pi\right]/(L+1)}\approx \epsilon(k_n'+i\kappa_L'),
\end{eqnarray}
provided that $e^\alpha\gg e^{-\alpha}$, where $k_n':=2n\pi/(L+1)$~\cite{SuppMat}.
Taking $\kappa_L$ and $\kappa_L'$ as functions of $\mu$, we have $\kappa_L(\mu)\approx \kappa_L'(e^{2\alpha}/\mu)$ for a sufficiently large system, suggesting a duality between $\mu=\mu_{\alpha}^{\pm}$ with $\mu_{\alpha}^{\pm}=e^\alpha A^{\pm1}$ parametrized by a variable $A$, with $\mu_{\alpha}^{+}=\mu_{\alpha}^{-}$ at $A=1$.

This duality can be seen in both the spectrum and the eigenstate accumulation, which can be characterized by the inverse participation ratio (IPR) defined as $I_n=\sum_{x}|\psi_{x,n}|^4$ for a given eigenstate. The IPR approaches $1$ for a perfectly localized state, and $1/(L+1)$ for a spatially homogeneous one. To further characterize the different directions of the SFA states, we define a directed IPR as $I_{d,n}=\sum_{x}(x_c-x)|\psi_{x,n}|^4/(L/2)$, with $x_c=L/2$ being the center of the system. By definition, $I_d$ takes positive (negative) values for states accumulating at $x=0$ ($x=L$), and $I_d=0$ for a spatially homogeneous state.

\begin{figure}
\includegraphics[width=1\linewidth]{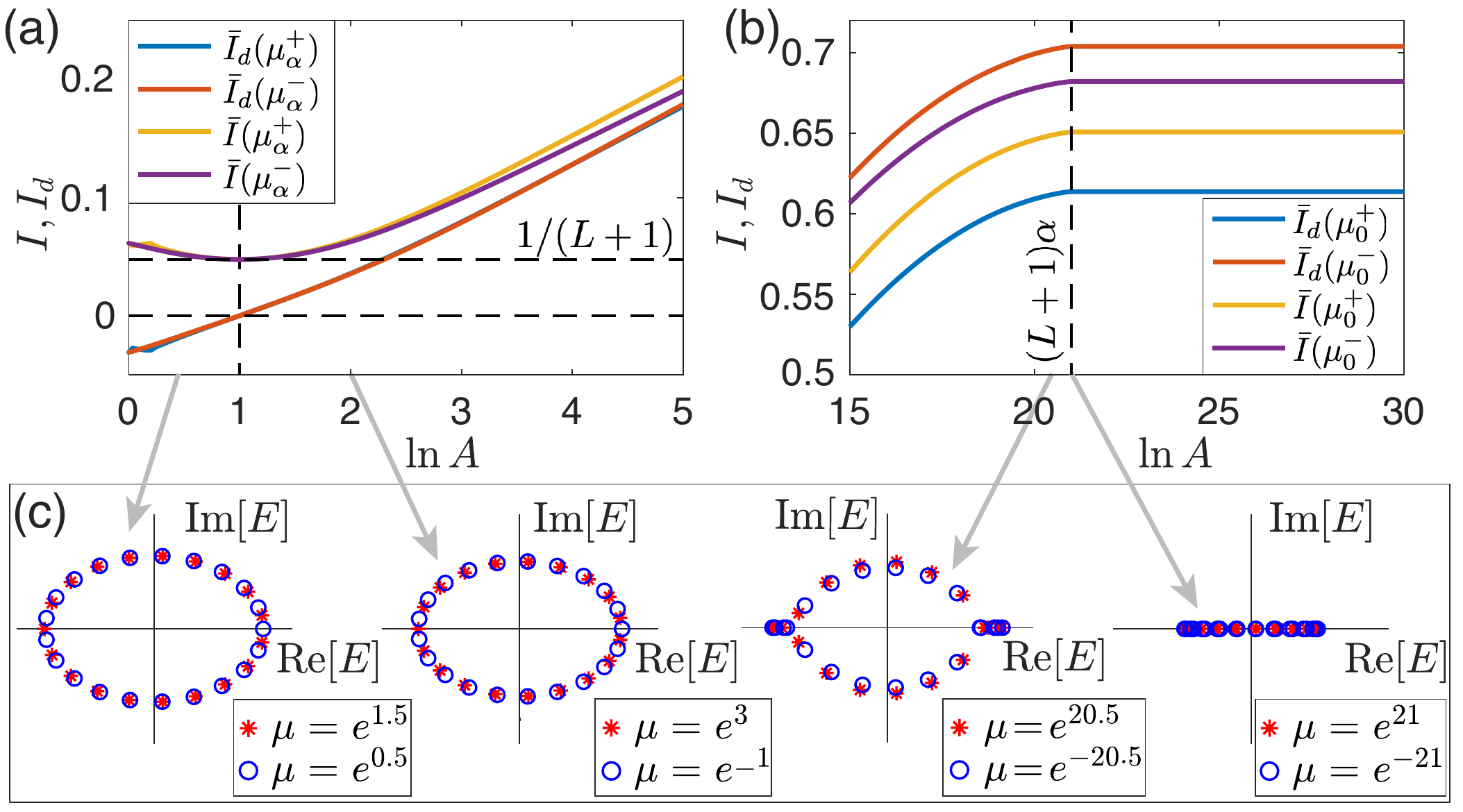}
\caption{(a) and (b) Average IPRs defined as $\bar{I}(\mu)=\sum_n I_n/N$ and $\bar{I}_d(\mu)=\sum_{n} I_{n,d}/N$. The summation of $n$ runs over all continuous eigenstates, and $N=L-1$ ($L+1$) is their total number in the presence (absence) of the pair of isolated states  $E_{\rm iso}\approx\pm\mu$. 
Colors of the curves indicate IPRs for different choices of $\mu$. 
Blue and red curves are almost identical in (a).
(c) Spectra with dual parameters, as indicated by the gray arrows. A phase transition to OBC-like line-spectrum at $\mu=e^{\pm(L+1)\alpha}=e^{\pm 21}$, with the parameters $L=20$ and $\alpha=1$.
}
\label{fig:duality}
\end{figure}

In Fig.~\ref{fig:duality}(a), we take averages over all continuous states for the IPRs and directed IPRs ($\bar{I}(\mu_\alpha^\pm)$ and $\bar{I}_d(\mu_\alpha^\pm)$), and present them as functions of $A$.
Note that for $\mu\gg 1$, the continuous eigenstates have vanishing amplitude at $x=0$, analogous to a system with $L$, not $L+1$ sites. Therefore, to properly compare the averaged IPRs between large and small $\mu$, 
they are rescaled as $(\bar{I},\bar{I}_d) \rightarrow (\bar{I},\bar{I}_d)L/(L+1)$ for $\mu>e^\alpha$, and the system's center is redefined as $x_c=(L-1)/2$ for the directed IPR.
We can see from Fig.~\ref{fig:duality}(a) that the quasi-PBCs and PBCs are recovered at $A=e^\alpha$ for
$\mu=\mu_{\alpha}^{\pm}$
respectively, where $\bar{I}(\mu)=1/(L+1)$ and $\bar{I}_d(\mu)=0$ as all eigenstates are fully delocalized. The IPR profiles agree well between the dual values of $\mu$ in the regime close to PBCs and quasi-PBCs ($A\sim e^\alpha$), but begin to diverge when $A$ gets larger.

To understand this divergence, we unveil a second duality between $\mu\sim e^{(L+1)\alpha}$ and $\mu\sim e^{-(L+1)\alpha}$, the latter corresponds to a transition between the qualitative spectral properties found for PBCs (loops) and OBCs (lines)~\cite{koch2020bulk}.
In Fig.~\ref{fig:duality}(b), we illustrate both IPRs for $\mu=\mu_0^{\pm}$ as functions of the variable $A$, i.e. $\mu_0^{\pm}=A^{\pm1}$. The above PBC-OBC transition 
is seen 
as $\bar{I}(\mu_0^{-})$ and $\bar{I}_d(\mu_0^{-})$ become constant for $\ln A\geqslant (L+1)\alpha$, reflecting the OBC skin modes. 
Interestingly, a similar transition also occurs at large $\mu=e^{(L+1)\alpha}$, characterized by the constant IPRs in Fig.~\ref{fig:duality}(b) when $\mu$ exceeds the critical value, indicating a second duality between $\mu=\mu_0^{\pm}$ in the large $A$ limit. 
These IPRs take different saturation values mainly because of the rescaling in the large $\mu$ regime with effectively different number of sites.
The critical value for this transition can also be identified from our approximation of Eq. (\ref{eq:dis_hopping}), where the decay exponent $\kappa_L=\alpha$ at $\mu=e^{(L+1)\alpha}$, 
recovering the decay exponent (and the imaginary flux) $\kappa_{\rm OBC}$ for NHSE under OBCs.
In Fig.~\ref{fig:duality}(c), we illustrate the spectra with several pairs of dual parameters, clearly showing the two types of dualities and the transition to a OBC-like spectrum.

\noindent{\it Co-existence of different regimes.--} 
The decay exponents $\kappa_L(\mu)$ of SFA states, as induced by the impurity, are insensitive to the exact configuration of non-reciprocal hoppings in the bulk. By contrast, skin modes under OBCs may have $k$-dependent decay exponents $\kappa_{\rm OBC}(k)$ if the system has hoppings beyond nearest neighbors~\cite{lee2019unraveling}.  
Requiring $\kappa_L(\mu_c)=\kappa_{\rm OBC}(k)$, one finds a $k$-dependent critical value of $\mu_c(k)$, with an intriguing consequence, namely, the co-existence of the SFA and NHSE for different eigenstates at a fixed $\mu$. Physically, this coexistence arises because at different wavenumbers $k$, an eigenstate effectively experiences couplings across different distances.

\begin{figure}
\includegraphics[width=1\linewidth]{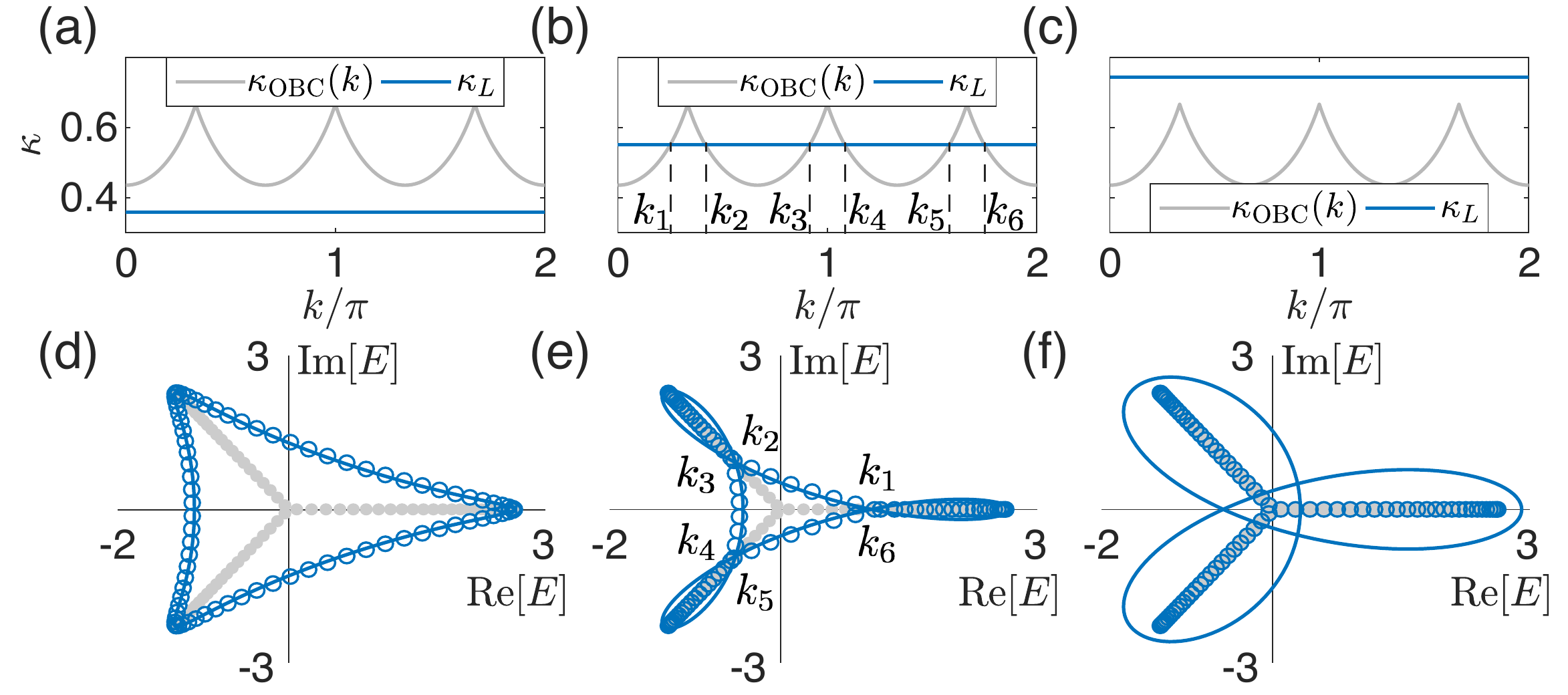}
\caption{(a)-(c) Decaying exponents $\kappa_L$ and $\kappa_{\rm OBC}(k)$ of the SFA and NHSE respectively,  for the Hamiltonian of Eq. (\ref{eq:H_NN}) with different boundary impurities $\mu$.
(d)-(f) Their corresponding spectra under different parameters and different boundary conditions. 
The blue lines, blue circles, and gray dots in (d)-(f) correspond to the spectra of the SFA where $E=\epsilon(k+i\kappa_L)$, numerical results with a boundary impurity, and the NHSE where $E=\epsilon(k+i\kappa_{\rm OBC}(k))$ respectively, with $\epsilon(k)$ the eigenenergy under PBCs. The parameters are $\alpha=1$, $L=80$, and $\mu=10^{-30}$, $10^{-45}$, and $10^{-60}$ from left to right. 
}
\label{fig:NN}
\end{figure}

Consider a system with different forward and backward couplings ranges and an impurity between $x=0$ and $L$:
\begin{equation}
H_{\rm NNN}=\sum_{x=0}^{L-1} e^{\alpha}\hat{c}^\dagger_x\hat{c}_{x+1}+\mu e^\alpha\hat{c}^\dagger_L\hat{c}_{0}+\sum_{x=0}^{L}e^{-\alpha}\hat{c}^\dagger_x\hat{c}_{x-2}.
\label{eq:H_NN}
\end{equation}
The decay exponents for the SFA at $\mu\ll 1$ and the NHSE at $\mu=0$ can be obtained as~\cite{SuppMat,lee2019unraveling}
\begin{eqnarray}
\kappa_L(\mu)=\frac{-\ln\mu}{L+1},~\kappa_{\rm OBC}(k)=\frac{1}{3}\ln\left[\frac{e^{2\alpha}}{2\cos(k-2j\pi/3)}\right],\nonumber
\end{eqnarray} 
with $j=\lfloor(k+\pi/3)/(2\pi/3)\rfloor$.
In Fig.~\ref{fig:NN}(a)-(c) we illustrate these two quantities versus $k$ for different  $\mu$. Together with the spectra in Fig.~\ref{fig:NN}(d)-(f), we see that an eigenstate always obeys the localization behavior with the smaller decaying exponent.
That is, all eigenstates exhibit the SFA when $\kappa_L<\kappa_{\rm OBC}(k)$ in Fig.~\ref{fig:NN}(a) and (d), and the NHSE when $\kappa_L>\kappa_{\rm OBC}(k)$ in Fig.~\ref{fig:NN}(c) and (f). In the intermediate regime of Fig.~\ref{fig:NN}(b) and (e), the SFA and NHSE co-exist for different $k$, as the spectrum follows the prediction of SFA when $k\in[k_{2m-1},k_{2m}]$ ($m=1,2,3$) 
where $\kappa_L$ is smaller, and the prediction of the NHSE otherwise, 
with $k_{2m}$ and $k_{2m-1}$ being the six special momentum values marked on Fig. \ref{fig:NN}(b) for which $\kappa_L=\kappa_{\rm OBC}$.
As also seen from Fig.~\ref{fig:NN}, due to the possibility of coexistence of the SFA and NHSE accumulation, even the qualitative spectral features are extremely sensitive to boundary impurity parameter $\mu$, an observation of general interest when it comes to build a sensing platform based on non-Hermiticity.

\noindent{\it Proposed experimental demonstration.--}
As steady-state phenomena, the SFA can be most easily demonstrated in an electrical circuit setting. In place of the Hamiltonian, we consider the circuit Laplacian $J$ which governs its steady state response via $\bold I =J\bold V$, where the components of $\bold V$ and $\bold I$ are respectively the electrical potentials and input currents at each node. The eigenspectra and eigenstates of $J$ can be directly resolved by measuring the voltage profile~\cite{lee2018topolectrical, helbig2019band} viz.
\begin{eqnarray}
V_\alpha =  J^{-1}_{\alpha\beta}I_\beta=\sum_{\lambda} \frac{ \langle\alpha|\psi^R_\lambda\rangle\langle\psi^L_\lambda|\bold I\rangle}{\epsilon_\lambda}
\end{eqnarray}
where $|\psi^{L/R}_\lambda\rangle$ are the left/right eigenvectors of $J$ corresponding to eigenvalue $\epsilon_\lambda$, and $V_\alpha$, $\langle\alpha|\psi^R_\lambda\rangle$ are respectively the potential and $\psi^R_\lambda$ values  at the $\alpha$-th node. 

To isolate a particular $\lambda'$-th eigenmode, we tune the circuit until $\epsilon_{\lambda'}\approx 0$, either by adjusting its variable components or by varying the AC frequency $\omega$~\cite{lee2018topolectrical}. $V_\alpha$ is then dominated by $\epsilon_{\lambda'}^{-1}\langle\alpha|\psi^R_{\lambda'}\rangle\langle\psi^L_{\lambda'}|\bold I\rangle$. If we further connect an input current $I_0$ to a fixed node $\beta'$ (the current leaves via the ground), $\langle\psi^L_{\lambda'}|\bold I\rangle=I_0\langle \beta'|\psi^L_{\lambda'}\rangle^*$ and the eigenstate profile $\langle\alpha|\psi^R_{\lambda'}\rangle$ across all nodes $\alpha$ becomes approximately proportional to the measured potential profile $V_\alpha$ i.e.
\begin{equation}
\psi^R_{\lambda'} \approx \frac{\epsilon_{\lambda'}\bold V}{I_0\langle \beta'|\psi^L_{\lambda'}\rangle^*}\propto \bold V.
\label{psiV}
\end{equation}
In other words, $\psi_{\lambda'}^R$ can be approximately measured through $\bold V$ when it is topolectrically resonant ($\epsilon_{\lambda'}\approx 0$).

\begin{figure}
	\includegraphics[width=\linewidth]{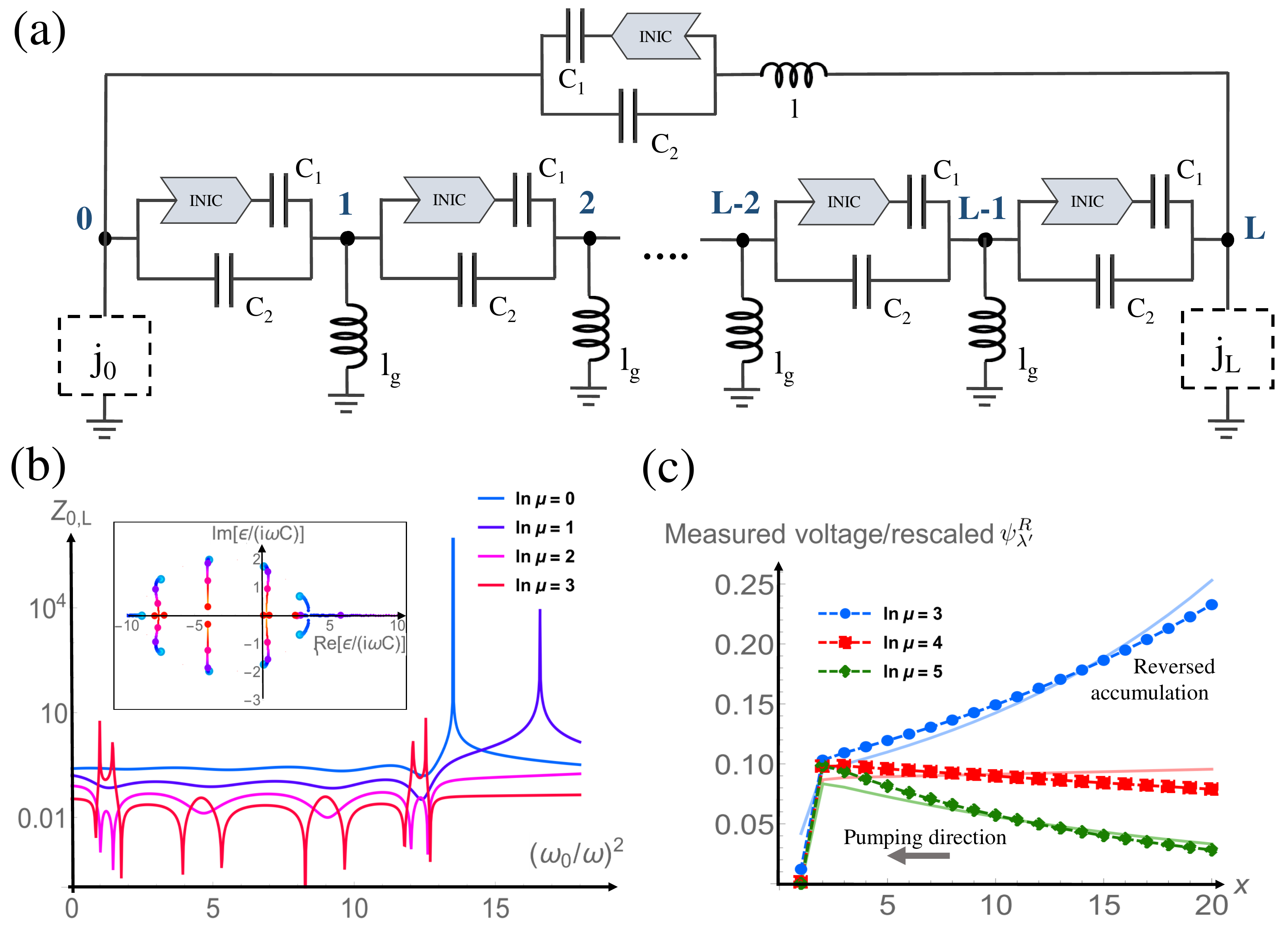}
\caption{(a) The circuit of Eq.~\ref{J}, whose asymmetric couplings are implemented through INICs and capacitors. The extra variable inductor $l$ gives rise to a coupling ``impurity''. Suitably designed grounding elements~\cite{SuppMat} enable desired eigenstates to be isolated at appropriate $\omega$. 
(b) Simulated impedance measurements across the impurity at various $\mu$, 
with position $\omega$ and height of impedance peaks loosely corresponding to the real and imaginary parts of the spectrum $\epsilon/(i\omega C)$ (inset). Parameters used are $L=9$, $C_1=1$ and $C_2=3$, so that $C=\sqrt{8}$ and $\alpha=0.364$. (c) Simulated electrical potential measurements vs. the profile of the bulk eigenstate with largest $\epsilon/(i\omega C)$, tuned close to resonance via Eq.~\ref{eq:E_hopping}. Not only is Eq.~\ref{psiV} accurate, the nature of eigenstate accumulation also agrees perfectly with the regimes of Fig.~1d. Parameters are $L=20$, $C_1=2.9$, $C_2=3$, such that $2\alpha\approx 4$.} 
\label{fig:circuit}
\end{figure}

A circuit Laplacian $J$ with a similar form as $H$ of Eq.~1 can be realized with the $L+1$-node LC circuit of Fig.~\ref{fig:circuit}a. Adjacent nodes acquire asymmetric non-Hermitian couplings through an INIC~\cite{hofmann2019chiral,helbig2020generalized} in series with a capacitor $C_1$, which together contribute an admittance of $i\omega C_1\left(\begin{matrix}-1 & 1 \\-1 & 1\end{matrix}\right)$ to the Laplacian~\cite{hofmann2019chiral,SuppMat}. The extent of asymmetry is regulated by another parallel capacitor $C_2$, such that $C_1/C_2=\tanh\alpha$~\cite{SuppMat}. To implement an ``impurity'' coupling between nodes $L$ and $0$, we connect an extra variable inductor $l$ in series with the parallel INIC $+$ capacitors configuration, such that the admittance in both directions is uniformly scaled by a factor of $\mu(\omega)= \mu(\omega)=(1-\omega^2 l(C_2-C_1))^{-1}$~\cite{SuppMat}.
To measure the profile of a desired eigenmode $\psi^R_{\lambda'}$, we first need to shift its eigenvalue $\epsilon_{\lambda'}$ maximally close to $0$. This can be achieved with additional identical grounding inductors $l_g$ at each bulk node, together with more carefully designed grounding circuits $j_0,j_L$ at the impurity nodes~\cite{SuppMat}. 
In all, our circuit Laplacian takes the form 
\begin{align}
J=& \biggl[\mu(\omega)\left(e^\alpha|L\rangle\langle 0|+e^{-\alpha}|0\rangle\langle L|\right)+\sum_{x=0,\pm}^{L-1}e^{\pm\alpha}|x\rangle\langle x\pm 1|
\notag\\
&-\sum_{x=0}^{L}\left(2 \cosh\alpha-\omega_0^2/\omega^2\right)|x\rangle\langle x|\biggl]\times i\omega C
\label{J}
\end{align}
where $C_1=C\sinh \alpha$, $C_2=C\cosh\alpha$ and $\omega^{-2}_0=l_{gr}C$, which is equal to $-i\omega C H$ (Eq.~1) up to a tunable real shift (the $|x\rangle\langle x$ term).  
Plotted in Fig.~\ref{fig:circuit}b are simulated impedance measurements $Z_{0,L}=\sum_\lambda\epsilon_\lambda^{-1}\langle \Delta|\psi^R_\lambda\rangle\langle\psi_\lambda^L|\Delta\rangle$, $|\Delta\rangle=|0\rangle-|L\rangle$~\cite{lee2018topolectrical} across the impurity as $\omega$ is varied, for different $\mu(\omega)=\mu$ adjusted through the inductors $l$. The impedance peaks correspond to values of $\omega$ where a Laplacian eigenvalue $\epsilon\approx 0$. 
For instance, the strongest peaks belonging to $\ln\mu=0,1$ arise from eigenvalues already on the real line (inset), while the weakest peaks from $\ln\mu=2$ are due to eigenvalues far from the real line. The entire spectrum (inset) can be reconstructed via systematic impedance measurements~\cite{helbig2019band,li2020critical}.

At these impedance peaks, the potential profile approximately corresponds to the eigenstate profile of the resonant eigenmode, as verified by simulated measurements [Fig~\ref{fig:circuit}c]. We clearly observe reversed and non-reversed eigenstates at different $\mu$, perfectly as predicted [Figs.~1d,2a]. Physically, the reversed voltage profile is a steady-state solution that represents a compromise between the competing non-reciprocal feedback mechanisms from the op-amps in the INICs. Scale-free behavior can be similarly detected when new nodes are introduced. More generally, we expect to measure these new forms of impurity-induced eigenstate accumulation in a variety of media whose steady-state description involve non-Hermitian asymmetric couplings~\cite{helbig2020generalized,hofmann2020reciprocal,weidemann2020efficient}.

{\it Discussion.-}
Boundary impurities in a non-Hermitian non-reciprocal lattice are found to induce rich transitions between NHSE, Bloch-like and SFA eigenstates along or against the direction of non-reciprocity, with stimulating duality relations between cases of weak and strong impurity strength.  Recognizing now that the well-known NHSE is only one of many impurity-induced consequences, a new basket of non-Hermitian phenomena may be explored, with the coexistence of SFA and NHSE shown as an example. 

\begin{acknowledgments}
\vspace{0.3cm}
\noindent{\textit{Acknowledgements.-}} 
J.G. acknowledges support from the Singapore NRF Grant No. NRF-NRFI2017-04 (WBS No. R-144-000-378-281). CH acknowledges support from the Singapore MOE Tier I grant (WBS No. R-144-000-435-133).
\end{acknowledgments}

\onecolumngrid
\begin{center}
\textbf{\large Supplementary Materials}\end{center}
\setcounter{equation}{0}
\setcounter{figure}{0}
\renewcommand{\theequation}{S\arabic{equation}}
\renewcommand{\thefigure}{S\arabic{figure}}
\renewcommand{\cite}[1]{\citep{#1}}

\section{Derivation of circuit Laplacian}

Here we provide a detailed derivation of the Laplacian (Eq.~11 of the main text) of the circuit as illustrated in Fig.~5 of the main text, and also furnish more details about its grounding connections. This circuit design is inspired by previous experimental cicuit realizations of various topological and non-Hermitian states~\cite{ningyuan2015time,lee2018topolectrical,imhof2018topolectrical,kotwal2019active,lu2019probing,olekhno2020topological,lee2019imaging,bao2019topoelectrical,zhang2020topolectrical}.

The Laplacian $J$ is defined as the operator that connects the vectors of input current and electrical potential via $\bold I = J\bold V$. In this work, we design a circuit array that (i) is non-Hermitian and non-reciprocal, with right/left couplings proportional to $e^{\pm \alpha}$, (ii) has special impurity couplings (in both directions) that are stronger than the rest by a tunable factor of $\mu=\mu(\omega)$ and (iii) also contains suitable grounding elements that allows the Laplacian eigenvalue spectrum to be shifted uniformly as desired. 

\begin{figure}
\includegraphics[width=.85\linewidth]{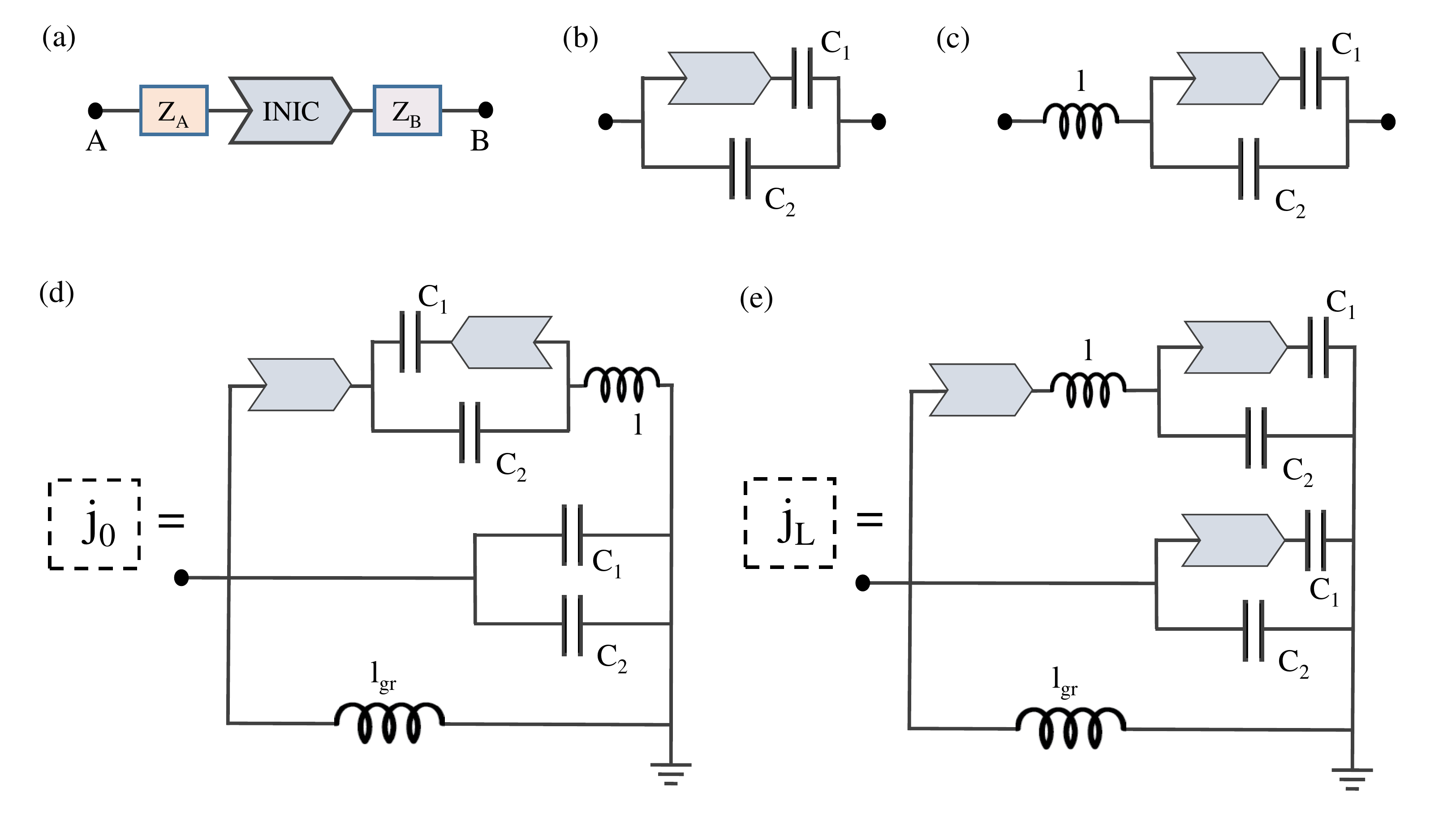}
\caption{(a) Two generic elements with impedances $Z_A,Z_B$ connected in series at either side of an INIC (elaborated in Ref.~\cite{hofmann2019chiral}) give rise to a non-Hermitian Laplacian Eq.~\ref{INIC}. (b) Asymmetric couplings of the simplest form (Eq.~\ref{JNN}) can be realized with a parallel configuration containing one INIC and two capacitors. (c) An impurity bond consisting of tunable equivalently rescaled asymmetric couplings (Eq.~\ref{Jimp}) can be realized with a variable inductor $l$ connected in series with $J_\text{NN}$. (d,e) Explicit example realizations of $j_0,j_L$ grounding components needed to make the grounding terms of the impurity nodes equivalent to the others'. Changing the AC frequency $\omega$ effects an uniform shift in Laplacian eigenvalues through $(i\omega l_{gr})^{-1}$.
}
\label{fig:circuitsupp}
\end{figure}

For (i), the unbalanced couplings $\propto e^{\pm \alpha}$ can be implemented by a parallel configuration of a capacitor $C_2$, and a combination of another capacitor $C_1$ that is connected in series with an INIC (negative impedance converter with current inversion). As elaborated in Ref.~\cite{hofmann2019chiral}, an INIC is an arrangement of operation amplifiers (op-amps) that reverses the sign of the impedance of components ``in front of'' it. Specifically, for a generic ideal INIC configuration as shown in Fig.~\ref{fig:circuitsupp}a, the input currents and potentials at the two ends obey
\begin{equation}
\left(\begin{matrix} I_A \\ I_B \end{matrix}\right)=\frac1{Z_A-Z_B}\left(\begin{matrix} 1 & -1 \\ 1 & -1 \end{matrix}\right)\left(\begin{matrix} V_A \\ V_B \end{matrix}\right)
\label{INIC}
\end{equation}
where $Z_A,Z_B$ are the impedances of components A and B. The Laplacian matrix above is not just asymmetric and hence non-Hermitian, but is also inversely proportional to the \emph{difference} between the two impedances, contrary to the usual case without the INIC.  

To implement the $\propto e^{\pm \alpha}$ couplings, we consider parallel configurations of two capacitors $C_1,C_2$, one on its own, and the other in series with an INIC (Fig.~\ref{fig:circuitsupp}b). This gives a Laplacian contribution of 
\begin{equation}
J_\text{NN}=i\omega C_1\left(\begin{matrix} 1 & -1 \\ 1 & -1 \end{matrix}\right)+ i\omega C_2\left(\begin{matrix} 1 & -1 \\ -1 & 1 \end{matrix}\right)= i\omega\left(\begin{matrix} C_2-C_1 & C_1-C_2 \\ -C_1-C_2 & C_1+C_2 \end{matrix}\right)= i\omega C\left(\begin{matrix} e^{-\alpha} & -e^{-\alpha} \\ -e^{\alpha} & e^{\alpha} \end{matrix}\right)
\label{JNN}
\end{equation}
if we set $C_1=C\sinh\alpha$, $C_2=C\cosh\alpha$, $C=\sqrt{C_2^2-C_1^2}$ a reference capacitance scale. If we connect each node of a OBC linear circuit array with these parallel configuration units, we end up with the Laplacian
\begin{align}
J_\text{OBC}&=i\omega \biggl[(C_2-C_1)|0\rangle\langle 0|+(C_2+C_1)|L\rangle\langle L|+\sum_{x=1}^{L-1}2C_2|x\rangle\langle x|-\left(\sum_{x=0}^{L-1}(C_2+C_1)|x\rangle\langle x+1|+(C_2-C_1)|x\rangle\langle x-1|\right)\biggl]\notag\\
&=i\omega C\biggl[e^{-\alpha}|0\rangle\langle 0|+e^\alpha|L\rangle\langle L|+\sum_{x=1}^{L-1}(2\cosh\alpha)|x\rangle\langle x|-\left(\sum_{x=0}^{L-1}e^\alpha|x\rangle\langle x+1|+e^{-\alpha}|x\rangle\langle x-1|\right)\biggl].
\end{align}
Note that the coefficient of $|x\rangle\langle x|$ merely sums out the total outgoing hopping amplitude.

To implement (ii) the impurity couplings that are equally asymmetric, but larger than the other couplings by a factor of $\mu$, we connect a tunable inductor $l$ with admittance $(i\omega l)^{-1}$ in series with the abovementioned parallel configuration (Fig.~\ref{fig:circuitsupp}c). Elementary applications of Kirchhoff's law gives us
\begin{equation}
J_\text{imp. NN}= i\omega\frac{(i\omega l)^{-1}}{(i\omega l)^{-1}+i\omega (C_2-C_1)}\left(\begin{matrix} C_2-C_1 & C_1-C_2 \\ -C_1-C_2 & C_1+C_2 \end{matrix}\right)= \frac1{1-\omega^2(C_2-C_1)l}J_\text{NN}=\mu(\omega)J_\text{NN}
\label{Jimp}
\end{equation}
which is proportional to $J_\text{NN}$ at the two nodes coupled by the impurity, up to a factor of $\mu(\omega)=\frac1{1-\omega^2(C_2-C_1)l}$. In other words, the impurity strength $\mu(\omega)$ can be adjusted both by changing the AC frequency $\omega$, or by tuning the inductor $l$ itself. Note that the upper-left term of $J_\text{bdry. NN}$ reduces to the simple result that the combined impedance of components connected in series is just the sum of their impedances.

The third important feature (iii), which is the implementation of grounding components that allow for a uniform shift in Laplacian eigenvalues, is more tricky. With ground connections given by $J_\text{gr}=\sum_{x=0}^Lj_x|x\rangle\langle x |$, the circuit Laplacian we have is given by (impurity is between the $L$-th and $0$-th nodes)
\begin{align}
J=J_\text{OBC}+J_\text{imp. NN}+J_\text{gr}=i\omega C &\biggl[\mu(\omega)\left(e^\alpha|0\rangle\langle 0|+e^{-\alpha}|L\rangle\langle L|-e^\alpha|L\rangle\langle 0|-e^{-\alpha}|0\rangle\langle L|\right)+e^{-\alpha}|0\rangle\langle 0|+e^\alpha|L\rangle\langle L|\notag\\
&+\sum_{x=1}^{L-1}(2\cosh\alpha)|x\rangle\langle x|-\left(\sum_{x=0}^{L-1}e^\alpha|x\rangle\langle x+1|+e^{-\alpha}|x\rangle\langle x-1|\right)\biggl]+\sum_{x=0}^Lj_x|x\rangle\langle x |.
\label{JJ}
\end{align}
Notably, the on-site terms are not even uniform. For identification with the Hatano-Nelson model with a single coupling impurity (Eq.~1 of the main text), we need to add adding grounding terms such that they are not just uniform but also tunable i.e. giving rise to a tunable multiple of the $(L+1)$-by-$(L+1)$ identity matrix. Since nodes $1$ through $L-1$ already have the same onsite coefficient of $2i\omega C \cosh\alpha$, we just need to ground them via identical inductors $l_\text{gr}$, such that $j_x=(i\omega l_\text{gr})^{-1}$ for $x=1,...,L-1$. The more tricky part is grounding nodes $0$ and $L$ with appropriate sets of components with combined admittance $j_0,j_L$ such that all onsite terms are equal. We first tidy up Eq.~\ref{JJ} such that the NN couplings, bulk groundings and impurity groundings are grouped together:
\begin{align}
J=&i\omega C \biggl[
-\mu(\omega)\left(e^\alpha|L\rangle\langle 0|+e^{-\alpha}|0\rangle\langle L|\right)-\left(\sum_{x=0}^{L-1}e^\alpha|x\rangle\langle x+1|+e^{-\alpha}|x\rangle\langle x-1|\right)
\biggl]+\sum_{x=1}^{L-1}\left(2i\omega C_2+(i\omega l_\text{gr})^{-1}\right)|x\rangle\langle x|\notag\\
&+\left[i\omega\mu(\omega)(C_2+C_1)+i\omega(C_2-C_1)+j_0\right]|0\rangle\langle 0|+\left[i\omega\mu(\omega)(C_2-C_1)+i\omega(C_2+C_1)+j_L\right]|L\rangle\langle L|.
\end{align}
For all onsite terms to be equal, we hence require that
\begin{align}
j_0=(i\omega l_\text{gr})^{-1}+i\omega (1-\mu(\omega))(C_2+C_1),\\
j_L=(i\omega l_\text{gr})^{-1}+i\omega (1-\mu(\omega))(C_2-C_1).
\end{align}
Recall from Eq.~\ref{Jimp} that $i\omega\mu(\omega)(C_2\pm C_1)$ are the admittances of the  $J_\text{imp. NN}$ configuration with respect to the ground. The remaining admittances $i\omega(C_2\pm C_1)$ can be realized by the configuration of $J_{NN}$ (Eq.~\ref{JNN}). As such, $j_0$ and $j_L$ can be realized by the configurations illustrated in Figs.~\ref{fig:circuitsupp}d and e.

All in all, our circuit Laplacian takes the form
\begin{align}
J=&i\omega C \biggl[
-\mu(\omega)\left(e^\alpha|L\rangle\langle 0|+e^{-\alpha}|0\rangle\langle L|\right)-\left(\sum_{x=0}^{L-1}e^\alpha|x\rangle\langle x+1|+e^{-\alpha}|x\rangle\langle x-1|\right)
\biggl]+\sum_{x=0}^{L}\left(2i\omega C_2+(i\omega l_\text{gr})^{-1}\right)|x\rangle\langle x|
\end{align}
whose realization is illustrated in Fig.~5 of the main text.

\section{Scale-free accumulation in the Hatano-Nelson model with a boundary impurity}
\subsection{Strong impurity}
We consider the following Hamiltonian
\begin{eqnarray}
H&=&\sum_{x=0}^{L-1} e^{\alpha}\hat{c}^\dagger_x\hat{c}_{x+1}+e^{-\alpha}\hat{c}^\dagger_x\hat{c}_{x-1}\nonumber\\
&&+\mu_+\hat{c}^\dagger_L\hat{c}_{0}+\mu_-\hat{c}^\dagger_0\hat{c}_{L}.\label{eq:H_supp}
\end{eqnarray}
Solving eigen-equation $H\Psi_n=E_n\Psi_n$ with $\Psi_n=\sum_x^L\psi_{x,n}\hat{c}^\dagger_x|0\rangle$ the $n$th eigenstate of the system, we obtain the following recursive conditions
\begin{eqnarray}
e^{\alpha}\psi_{x+1,n}+e^{-\alpha}\psi_{x-1,n}&=&E_n\psi_{x,n}\label{eq:bulk_supp}
\end{eqnarray}
for $x=1,2,...,L-1$, and
\begin{eqnarray}
\mu e^{\alpha}\psi_{0,n}+e^{-\alpha}\psi_{L-1,n}&=&E_n\psi_{L,n},\label{eq:boundary1_supp}\\
e^{\alpha}\psi_{1,n}+\mu e^{-\alpha}\psi_{L,n}&=&E_n\psi_{0,n}.\label{eq:boundary2_supp}
\end{eqnarray}
Intuitively, when $\mu$ is large, two isolated solutions localized around $x=0$ and $x=L$ are expected due to the strong couplings between these two sites. 
Assuming these solutions decay exponentially from the two sites into the bulk, we find that they can be explicitly expressed as
\begin{eqnarray}
\psi_{0}^{+}=\psi_L^{+} e^{-\alpha},~\psi_{x}^{+}=\frac{e^{-\alpha}}{\mu}\psi_{x-1}^{+}~{\rm for}~x=1,2,...,
~\psi_{x}^{+}=\frac{e^{\alpha}}{\mu}\psi_{x+1}^{+}~{\rm for}~x=L,L-1,...;\\
\psi_{0}^{-}=-\psi_L^{-} e^{-\alpha},~\psi_{x}^{-}=-\frac{e^{-\alpha}}{\mu}\psi_{x-1}^{-}~{\rm for}~x=1,2,...,
~\psi_{x}^{-}=-\frac{e^{\alpha}}{\mu}\psi_{x+1}^{-}~{\rm for}~x=L,L-1,...,
\end{eqnarray}
whose eigenenergies are given by
\begin{eqnarray}
E^{\pm}_{\rm iso}=\pm(\mu+\frac{1}{\mu}).
\end{eqnarray}
These solutions are valid providing $\mu>e^{\pm\alpha}$, so that they indeed decay from $x=0$ and $x=L$ into the bulk;
and $(e^{\pm\alpha}/\mu)^L\sim 0$, so that they have vanishing amplitudes in the middle of the system. In the main text, we have assumed $\mu\gg e^{\pm\alpha}$, therefore the above conditions are satisfied and we have $E^{\pm}_{\rm iso}\approx\pm\mu$.

For convenience, we label these two isolated eigenstates with $n=0$ and $n=L$ respectively.
The other $L-1$ eigenstates of $n\in[1,L-1]$, referred as continuous eigenstates as they have a continuous spectrum, shall mainly distribute within the rest $L-1$ sites of the system with eigenenergies $E_n\ll\mu$, thus we shall have a vanishing $\psi_{0,n}$ from Eq. (\ref{eq:boundary1_supp}). We further consider an ansatz of exponentially decaying eigenstates given by 
\begin{eqnarray}
\psi_{0,n}\ll1,~\psi_{x,n}=e^{-M_n (x-1)}~{\rm for}~x\neq0.\label{eq:ansatz_supp}
\end{eqnarray}
Substituting the ansatz into Eq. (\ref{eq:boundary2_supp}), we obtain
\begin{eqnarray}
-e^{\alpha}=\mu e^{-\alpha} e^{-M_n(L-1)},\nonumber
\end{eqnarray}
yielding
\begin{eqnarray}
M_n=\frac{\ln\mu-2\alpha-i(2n+1)\pi}{L-1}:=\kappa_L-\frac{i(2n+1)\pi}{L-1}.\label{eq:exponent_supp}
\end{eqnarray}
However, Eqs. (\ref{eq:bulk_supp}) and (\ref{eq:boundary1_supp}) give different eigenenergies with this exponentially decaying solution.
A consistent solution can be obtained by further requiring $e^\alpha\gg e^{-\alpha}$ and $e^{\alpha}e^{-M_n}\gg e^{-\alpha}e^{M_n}$. The first condition corresponds to a strong non-reciprocity of the system, and the second one is equivalent to $\mu\ll e^{(L+1)\alpha}$, which is generally satisfied for a large enough system.
Under these conditions, Eq. (\ref{eq:bulk_supp}) gives
\begin{eqnarray}
E_n\approx e^\alpha e^{-M_n}= e^{\alpha}e^{- \left[\ln \mu -2\alpha-i(2n+1)\pi\right]/(L-1)}\approx\epsilon(k_n+i\kappa_L),
\end{eqnarray}
with $k_n:=(2n+1)\pi/(L-1)$, $n=1,2,...,L-1$, and $\epsilon(k)\approx e^{\alpha}e^{ik}$ the eigenenergies under PBCs and the strong non-reciprocity.
On the other hand, since now we have $\mu\gg e^\alpha\sim E_n\gg e^{-\alpha}$, the second term of Eq. (\ref{eq:boundary1_supp}) can be neglected, yielding
\begin{eqnarray}
|\psi_{0,n}|\approx |\frac{e^{-M_n L}}{\mu}|\approx\frac{e^{2\alpha}}{2\mu}\ll 1,
\end{eqnarray}
in consistent with the vanishing $\psi_{0,n}$ obtained previously.

\begin{figure*}
\includegraphics[width=1\linewidth]{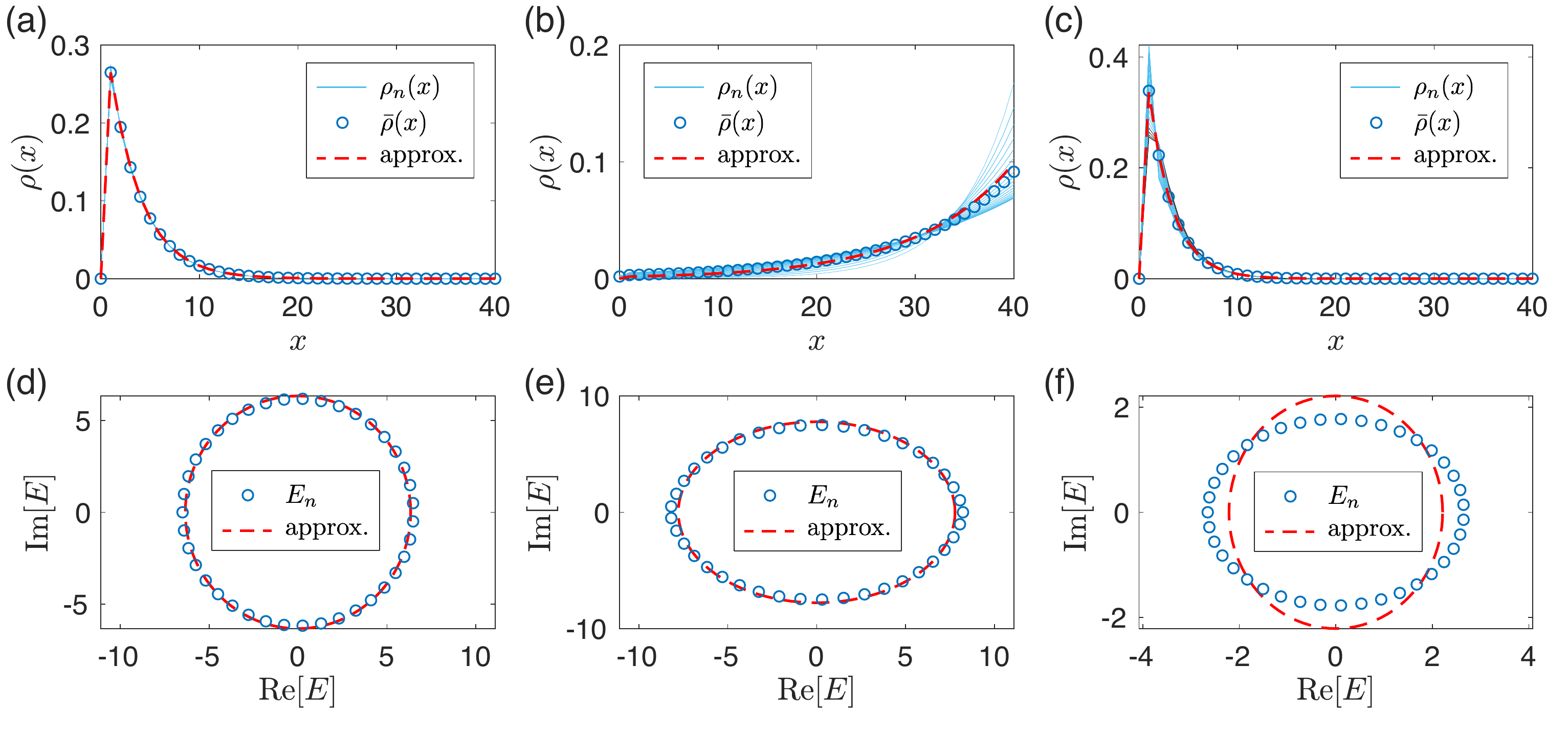}
\caption{(a)-(c) Spatial distribution of continuous eigenstates and their (d)-(f) eigenenergies of the model of Eq. (\ref{eq:H_supp}). The distribution is defined as $\rho_n(x)=|\psi_{x,n}|^2$ and $\bar{\rho}(x)=\sum_n \rho_n(x)/(L-1)$. Red dashed lines indicate analytical results from the approximation. Parameters are $L=40$, and (a)(d) $\alpha=2$, $\mu=e^10$; (b)(e) $\alpha=2$, $\mu=e^2$; and (c)(f) $\alpha=1$, $\mu=e^10$.}
\label{fig:accuracy}
\end{figure*}

In Fig.~\ref{fig:accuracy}, we compare numerical results under several different parameter regimes with the above approximation, which works well when $\mu\gg e^{\alpha}$, and $e^{\alpha}\gg e^{-\alpha}$, and $\mu\ll e^{(L+1)\alpha}$ [Fig.~\ref{fig:accuracy}(a) and (d)]. In most other parameter regimes, while the eigenenergies and distribution of individual eigenstates cannot be predicted accurately, the average distribution of all continuous eigenstates is still in good consistence with Eqs. (\ref{eq:ansatz_supp}) and (\ref{eq:exponent_supp}), as shown in the middle and right columns of Fig.~\ref{fig:accuracy}.
On the other hand, when $\mu$ approach the value of $e^{(L+1)\alpha}$, the system goes into the OBCs-like regime with the NHSE, and our approximation of the SFA is no longer valid, as discussed in the main text.

\subsection{Weak impurity}
Next we consider a weak impurity limit with $\mu\ll 1$ and a strong non-reciprocity $e^\alpha\gg e^{-\alpha}$, and another ansatz
\begin{eqnarray}
\psi'_{x,n}=e^{-M'_n x}.\label{eq:ansatz2_supp}
\end{eqnarray}
with $M_n>0$, as we observe no reversed accumulation in the regime with $\mu<1$. Thus Eq. (\ref{eq:boundary2_supp}) is simplified as 
\begin{eqnarray}
e^{\alpha}	e^{-M'_n}=E'_n.\label{eq:E_weak_dis_supp}
\end{eqnarray}
Substituting the above equation to Eq. (\ref{eq:boundary1_supp}) with its second term being neglected, one can obtain
\begin{eqnarray}
M'_n=\frac{-\ln \mu-i2n\pi}{L+1}:=\kappa_L'-\frac{i2n\pi}{L+1}.
\end{eqnarray}
This solution also confirms that the second term of Eq. (\ref{eq:boundary1_supp}) is neglectable comparing to the rest two terms.
The eigenenergies are thus directly given by Eq. (\ref{eq:E_weak_dis_supp}).

\section{Quasi-PBC delocalized eigenstates}
To gain further insights into the quasi-PBCs at $\mu=2\alpha$, let us exploit the following effective translational invariant Hamiltonian,
$\bar{H}_{\rm PBC}(k)=H_{\rm PBC}(k+i\kappa_L)$, with its real-space form reads as 
\begin{eqnarray}
\bar{H}_{\rm PBC}&=&\sum_{x=0}^{L} e^{-\kappa_L}e^{\alpha}\hat{c}^\dagger_x\hat{c}_{x+1}+e^{\kappa_L}e^{-\alpha}\hat{c}^\dagger_x\hat{c}_{x-1}.
\end{eqnarray}
In above site $L+1$ is understood as site 0. According to our spectral results in Eq.~\ref{eq:E_hopping} in the main text, $\bar{H}_{\rm PBC}$, though having an extra $\kappa_L$ related imaginary flux,  yields the approximate eigenvalues of our lattice system for $\mu=e^{(L-1)\kappa_L+2\alpha}$.
We next remove the imaginary flux in the bulk by applying a similarity transformation $\bar{H}_{\rm PBC}'=S_L^{-1}\bar{H}_{\rm PBC}S_L$ with $S_L={\rm Diag}\{1,e^{\kappa_L},e^{2\kappa_L},...,e^{\kappa_LL} \}$. This gives (without changing the eigenvalues)
\begin{eqnarray}
\bar{H}_{\rm PBC}'&=&\sum_{x=0}^{L-1} e^{\alpha}\hat{c}^\dagger_x\hat{c}_{x+1}+e^{-\alpha}\hat{c}^\dagger_x\hat{c}_{x-1}\nonumber\\
&&+e^{-\kappa_L(L+1)} e^{\alpha}\hat{c}^\dagger_L\hat{c}_{0}+e^{\kappa_L(L+1)} e^{-\alpha}\hat{c}^\dagger_0\hat{c}_{L}.
\end{eqnarray}
So long as $L$ is sufficently large, we still have $\kappa_L(L+1)\approx \ln\mu-2\alpha$, thus the boundary hopping in $\bar{H}_{\rm PBC}'$ shown above becomes
\begin{eqnarray}
 \mu^{-1}e^{3\alpha}\hat{c}^\dagger_L\hat{c}_{0}+\mu e^{-3\alpha}\hat{c}^\dagger_0\hat{c}_{L}.\label{eq:kappa_boundary}
\end{eqnarray}
It is seen that at $\mu=e^{2\alpha}$, $\bar{H}_{\rm PBC}'$ recovers the original Hamiltonian under PBCs.
This is fully consistent with the observation from Eq.~\ref{eq:dis_hopping} in the main text, namely, the decay exponent $\kappa_L=0$ for $\mu=e^{2\alpha}$. The above treatment is however more stimulating to digest situations with $\mu\neq e^{2\alpha}$, where the translational invariance of $\bar{H}_{\rm PBC}'$ is broken at the boundary. For $\mu>e^{2\alpha}$, the hopping from $x=L$ to $x=0$ is further enhanced whereas the opposite hopping is further suppressed  (as respectively compared with the translational invariant case).  The eigenstates are then expected to populate more at $x=0$. Likewise, eigenstates should accumulate more at $x=L$ when $\mu<e^{2\alpha}$, thereby exhibiting the reversed SFA.

\section{Different accumulating behaviors of the model with two non-reciprocity length scales}
We consider a system with nearest-neighbor backward couplings and next-nearest-neighbor forward couplings, and a local impurity between sites $x = 0$ and $x = L$, described by the Hamiltonian
\begin{eqnarray}
H_{\rm NNN}&=&\sum_{x=0}^{L-1} \left[e^{\alpha}\hat{c}^\dagger_x\hat{c}_{x+1}+\mu e^\alpha\hat{c}^\dagger_L\hat{c}_{0}\right]+\sum_{x=0}^{L}e^{-\alpha}\hat{c}^\dagger_x\hat{c}_{x-2}.
\label{eq:H_NN_supp}
\end{eqnarray} 
Solve eigen-function $H_{\rm NNN}\Psi_n=E_n\Psi_n$ with $\Psi_n=\sum_x^L\psi_{x,n}\hat{c}^\dagger_x|0\rangle$, the recursive conditions of $\psi_{x,n}$ are given by
\begin{eqnarray}
e^{\alpha}\psi_{x+1,n}+e^{-\alpha}\psi_{x-2,n}=E_n\psi_{x,n}\label{eq:bulk_NN_supp}
\end{eqnarray}
for $x=1,2,...,L-1$, and
\begin{eqnarray}
\mu e^{\alpha}\psi_{0,n}+e^{-\alpha}\psi_{L-2,n}&=&E_n\psi_{L,n},\label{eq:boundary1_NN_supp}\\
e^{\alpha}\psi_{1,n}+e^{-\alpha}\psi_{L-1,n}&=&E_n\psi_{0,n}.\label{eq:boundary2_NN_supp}
\end{eqnarray}
Similarly to the model of Eq. (\ref{eq:H_supp}) at weak impurity limit, we consider the parameter regime with $\mu\ll1$ and $e^\alpha\gg e^{-\alpha}$, and the same SFA solution can be obtained as 
\begin{eqnarray}
\psi_{x,n}=e^{-M_n x},~M_n=\frac{-\ln \mu-i2n\pi}{L+1}:=\kappa_L-\frac{i2n\pi}{L+1}.
\end{eqnarray}
On the other hand, to solve the OBC system with $\mu=0$, we consider the an imaginary flux $\kappa_{\rm OBC}(k)$ under PBCs, corresponding an effective Hamiltonian 
\begin{eqnarray}
\bar{H}_{\rm NNN}(k)=H_{\rm NNN}(k+i\kappa_{\rm OBC}(k))=ze^\alpha+e^{-\alpha}/z^2,
\end{eqnarray} 
with $z=e^{i[k+i\kappa_{\rm OBC}(k)]}$. The OBC system is described by a GBZ, where the eigenenergies satisfy $\bar{E}(k_1)=\bar{E}(k_2)$ for pairs of quasi-momenta with $\kappa_{\rm OBC}(k_1)=\kappa_{\rm OBC}(k_2)$. Numerically, we find that this condition is satisfied when $k_1+k_2=0$, $2\pi/3$, and $4\pi/3$, for $k_1,k_2\in [-\pi/3,\pi/3]$, $[\pi/3,\pi]$, and $[\pi,5\pi/3]$, respectively.
With these relations between $k_1$ and $k_2$, we obtain
\begin{eqnarray}
\kappa_{\rm OBC}(k)&=&\frac{1}{3}\ln\left[\frac{e^{2\alpha}}{2\cos(k-2j\pi/3)}\right],
\end{eqnarray} 
with $j=\lfloor(k+\pi/3)/(2\pi/3)\rfloor$.

\section{Co-existence of SFA and topological localization}
\begin{figure*}
\includegraphics[width=1\linewidth]{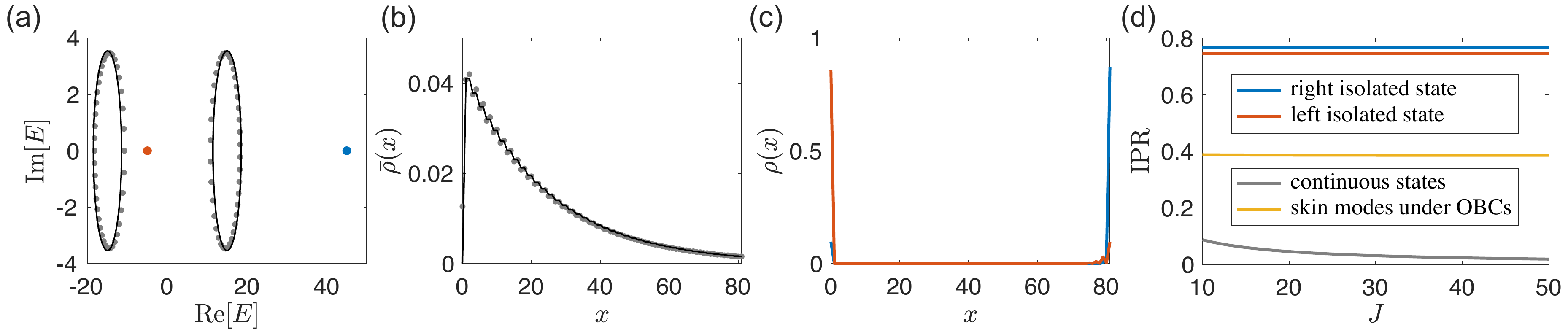}
\caption{(a) the full spectrum of the SSH model with a local on-site potential, gray dots are the continuous states, blue and red dots are the two topological/isolated states. (b) the average distribution of the continuous states. (c) The distributions of the topological/isolated states. 
(4d) IPR versus system's size for different eigenstates of the system with a local on-site potential, and for skin modes under OBCs. 
The results for continuous states and skin modes are the average values of all corresponding eigenstates. 
Parameters are $\alpha=2$, $t_{ab}=15$, $\mu=40$, and $J=40$ for (a,b,c).}
\label{fig:SSH}
\end{figure*}

In this section we consider a non-Hermitian topological system with non-reciprocal couplings, where SFA and topological localization exist for different eigenstates of the system. The explicit model we consider is a non-reciprocal Su-Schrieffer-Heeger (SSH) model \cite{SSH}, described by the Hamiltonian
\begin{eqnarray}
H&=&\sum_{j=0}^J e^{\alpha}\hat{a}^\dagger_j\hat{b}_j+e^{-\alpha}\hat{b}^\dagger_j\hat{a}_j\nonumber\\
&&+t_{ab}\hat{b}^\dagger_j\hat{a}_{j+1}+h.c.+\mu\hat{b}^\dagger_J\hat{b}_J
\end{eqnarray}
with $J+1$ the number of unit cells. Note that instead of enhanced boundary couplings, here spatial inhomogeneity is induced by an on-site potential acting on a single lattice site, so that only one isolated state shall emerge due to the impurity, in the absence of a nontrivial topology.
The existence of isolated boundary states and their connection to the bulk topology has been studied in Ref.~\cite{liu2020diagnosis}, and here we shall focus on the topologically nontrivial regime of $t_{ab}>e^\alpha$, with strong local potential $\mu\gg t_{ab}$.
By solving the eigen-equation $H\Psi_n=E_n\Psi_n$ with eigenstates defined as $\Psi_n\sum_j^J(\psi^a_{j,n}\hat{a}^\dagger_{j,n}+\psi^b_{j,n}\hat{b}^\dagger_{j,n})|0\rangle$,
we obtain the solutions 
\begin{eqnarray}
\psi^b_{j,n}&=&e^{-\left[\kappa_{J}-i(2n+1)\pi/J\right] j},\psi_{j,n}^a=\phi_{j-1,n}^b,\nonumber\\
\kappa_{J}&\approx&\left[\ln\mu -\alpha\right]/J\label{eq:decay_SSH1}
\end{eqnarray}
with a continuous spectrum
\begin{eqnarray}
E_n\approx \pm\left(t_{ab}+\frac{e^{\alpha-\kappa_J} e^{\left[\alpha-i(2n+1)\pi  \right]/J}}{2}\right),
\end{eqnarray}
exhibiting the same scale-free decaying behavior as in the main text, due to the $1/J$ coefficient in $\kappa_J$.

In the parameter regime we choose, the system holds two eigenstates isolated from the continuous spectrum, as shown in Fig.~\ref{fig:SSH}(a). 
Associated with the nontrivial bulk topology, these two states localized at $j=0$ and $j=J$ respectively [Fig.~\ref{fig:SSH}(b)], with the later one affected more by the local potential at $j=J$ and having an eigenenergy $E\approx \mu$. On the other hand, both of these two states exhibit a much stronger accumulation to the boundary, in contrast with the continuous states illustrated in Fig.~\ref{fig:SSH}(c).

To further characterize their difference, we calculate the inverse participation ratio (IPR) defined as ${\rm IPR}=\sum_{j}(|\psi^a_{j}|^4+|\psi^b_{j}|^4)$ for the isolated states, continuous states, and skin modes under OBCs (with $\mu=0$), and demonstrate it versus the system's size $J$ in Fig.~\ref{fig:SSH}(d). 
Besides their weaker accumulation reflected by a smaller IPR, the continuous states are less localized for a larger size of the system, due to the $1/J$ coefficient in the decaying exponent $\kappa_{J}$.
On the other hand, the IPR for isolated states, and for the skin modes under OBCs, remains a constant when increasing the size.

%
\end{document}